\documentclass[%
 reprint,
superscriptaddress,
 amsmath,amssymb,
 aps,
prb,
floatfix,
]{revtex4-2}

\usepackage{graphicx}
\usepackage{dcolumn}
\usepackage{bm}
\usepackage[version=3]{mhchem} 
\usepackage[T1]{fontenc}
\usepackage{siunitx}
\usepackage[T1]{fontenc}
\usepackage{siunitx}
\usepackage[export]{adjustbox}
\usepackage{hyperref}

\begin{document}


\title{Enhanced quantum magnetometry with a laser-written integrated photonic diamond chip}

\author{Yanzhao Guo}
\email{GuoY65@cardiff.ac.uk}
\affiliation{School of Engineering, Cardiff University, Queen’s Buildings, The Parade, Cardiff, CF24 3AA, United Kingdom}
\affiliation{Translational Research Hub, Cardiff University, Maindy Road, Cardiff, CF24 4HQ, United Kingdom}
\author{Giulio Coccia}
\affiliation{Department of Physics, Politecnico di Milano, Piazza Leonardo da Vinci, 32, 20133 Milano, Italy}
\affiliation{Institute for Photonics and Nanotechnologies (CNR-IFN), Piazza Leonardo da Vinci, 32, 20133 Milano, Italy}
\author{Vinaya Kumar Kavatamane}
\affiliation{Institute for Quantum Science and Technology, University of Calgary, Calgary, AB T2N 1N4, Canada}
\author{Argyro N. Giakoumaki}
\affiliation{Department of Physics, Politecnico di Milano, Piazza Leonardo da Vinci, 32, 20133 Milano, Italy}
\affiliation{Institute for Photonics and Nanotechnologies (CNR-IFN), Piazza Leonardo da Vinci, 32, 20133 Milano, Italy}
\author{Anton N. Vetlugin}
\affiliation{Centre for Disruptive Photonic Technologies, TPI, Nanyang Technological University, Singapore}
\affiliation{Division of Physics and Applied Physics, SPMS, Nanyang Technological University, Singapore}
\author{Roberta Ramponi}
\affiliation{Department of Physics, Politecnico di Milano, Piazza Leonardo da Vinci, 32, 20133 Milano, Italy}
\affiliation{Institute for Photonics and Nanotechnologies (CNR-IFN), Piazza Leonardo da Vinci, 32, 20133 Milano, Italy}
\author{Cesare Soci}
\affiliation{Centre for Disruptive Photonic Technologies, TPI, Nanyang Technological University, Singapore}
\affiliation{Division of Physics and Applied Physics, SPMS, Nanyang Technological University, Singapore}
\author{Paul E. Barclay}
\affiliation{Institute for Quantum Science and Technology, University of Calgary, Calgary, AB T2N 1N4, Canada}
\author{John P. Hadden}
 \affiliation{School of Engineering, Cardiff University, Queen’s Buildings, The Parade, Cardiff, CF24 3AA, United Kingdom}
\affiliation{Translational Research Hub, Cardiff University, Maindy Road, Cardiff, CF24 4HQ, United Kingdom}
\author{Anthony J. Bennett}
\email{BennettA19@cardiff.ac.uk}
 \affiliation{School of Engineering, Cardiff University, Queen’s Buildings, The Parade, Cardiff, CF24 3AA, United Kingdom}
\affiliation{Translational Research Hub, Cardiff University, Maindy Road, Cardiff, CF24 4HQ, United Kingdom}
\author{Shane M. Eaton}
\affiliation{Department of Physics, Politecnico di Milano, Piazza Leonardo da Vinci, 32, 20133 Milano, Italy}
\affiliation{Institute for Photonics and Nanotechnologies (CNR-IFN), Piazza Leonardo da Vinci, 32, 20133 Milano, Italy}

\date{\today}

\begin{abstract}
An ensemble of negatively charged nitrogen-vacancy centers in diamond can act as a precise quantum sensor even under ambient conditions. In particular, to optimize thier sensitivity, it is crucial to increase the number of spins sampled and maximize their coupling to the detection system, without degrading their spin properties. In this paper, we demonstrate enhanced quantum magnetometry via a high-quality buried laser-written waveguide in diamond with a 4.5 ppm density of nitrogen-vacancy centers. We show that the waveguide-coupled nitrogen-vacancy centers exhibit comparable spin coherence properties as that of nitrogen-vacancy centers in pristine diamond using time-domain optically detected magnetic resonance spectroscopy. Waveguide-enhanced magnetic field sensing is demonstrated in a fiber-coupled integrated photonic chip, where probing an increased volume of high-density spins results in \SI{63}{\pico T\cdot\hertz^{-1/2}} of DC-magnetic field sensitivity of and \SI{20}{\pico T\cdot\hertz^{-1/2}} of AC magnetic field sensitivity. This on-chip sensor realizes at least an order of magnitude improvement in sensitivity compared to the conventional confocal detection setup, paving the way for microscale sensing with nitrogen-vacancy ensembles.
\end{abstract}
\maketitle


\section{\label{Introduction}Introduction}

Quantum sensing with negatively charged nitrogen-vacancy centers (NVs) in diamond has attracted broad interest in the last two decades\cite{Balasubramanian2008NanoscaleConditions, Barry2020SensitivityMagnetometry}. Due to its asymmetric atomic structure\cite{Maze2011PropertiesApproach}, the NV is highly sensitive to weak external influences like temperature\cite{Acosta2010TemperatureDiamond}, pressure\cite{Doherty2014ElectronicPressure, Mittiga2018ImagingDiamond}, electric field\cite{Dolde2011Electric-fieldSpins} and magnetic field \cite{Rondin2014MagnetometryDiamond} at the nanoscale\cite{Balasubramanian2008NanoscaleConditions}. Meanwhile, their long spin coherence time \cite{Balasubramanian2009UltralongDiamond} allows NV-based quantum sensors to achieve remarkable sensitivity\cite{Barry2020SensitivityMagnetometry}. However, their relatively low optical excitation efficiency and finite photon collection efficiency have limited their sensitivity in practice\cite{Barry2020SensitivityMagnetometry,Doherty2013TheDiamond}. Although various submicron-integrated photonic structures for single NV have been studied to enhance the photon collection rate\cite{Pachlatko2024NanoscaleMicrostrip.,Wang2022Self-alignedPrecision}, few can easily be applied to the ensemble of NVs, and most would degrade the coherence properties of NVs. In particular,  given the sensitivity scales inversely with the square root of detected signal intensity\cite{Barry2020SensitivityMagnetometry, Rondin2014MagnetometryDiamond}, it is crucial to efficiently excite and collect from a large volume of ensemble NVs with good spin coherence properties.

Previous work on NV ensemble sensing was based on the millimeter size of diamond devices with sub-nanotesla sensitivities. For example, a light-trapping diamond waveguide geometry\cite{Clevenson2015BroadbandWaveguide} improves the probed number of  NVs via increasing optical depth in the mm-sized bulk diamond, or the integration of optical fiber tip with a millimeter size of diamond\cite{PhysRevApplied.14.044058}. There are few studies on the scaling of these sensors towards the micron scale\cite{PhysRevX.5.041001}. One solution could be to couple an ensemble of NVs into a fiber-integrated photonic structure, with confined micron-scale sensing resolution, for example, the waveguide integrated NVs in diamond\cite{Hoese2021IntegratedArrays,Hadden2018IntegratedWriting}. Recently, laser writing has been demonstrated as a powerful tool for fabricating micron photonic circuits with integrated quantum emitters from single to ensemble level\cite{Eaton2019QuantumIrradiation,Guo2024Laser-writtenDiamond}. Moreover, laser-written waveguide-integrated NVs (WGINVs ) have been shown to have comparable spin coherence properties to native NVs in diamond\cite{Guo2024Laser-writtenDiamond}. 

    \begin{figure*}[ht]
\centering
\includegraphics[width=\textwidth]{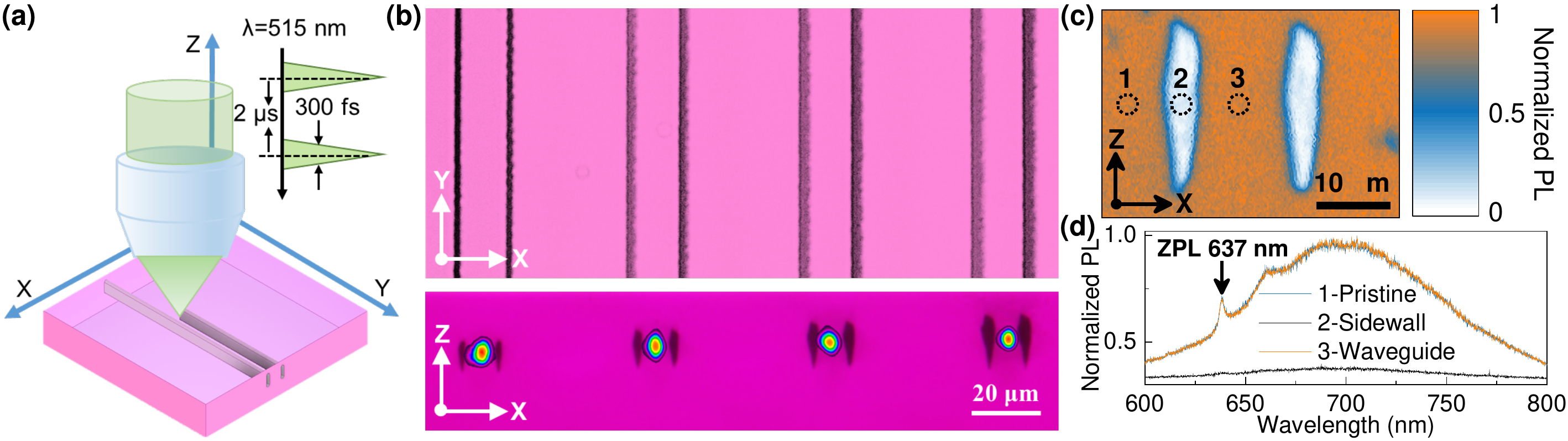}
\caption{(a) Schematic diagram for the laser writing waveguides in DNV-B14 diamond. (b) Overhead (upper plate) and cross-sectional (lower plate) optical microscopy images of waveguides with overlaid 635 nm mode profiles. (c) The confocal PL map of the WGINVs was written with a 40 mW femtosecond laser, mapped from the y-direction (along the waveguide). (d) PL emission spectrum for the 1-pristine, 2-waveguide, and 3-sidewall region in waveguide written with 40 mW.}
\label{Fig1}
\end{figure*}

In this paper, we demonstrate an on-chip micron sensor with enhanced sensitivity via a laser-written waveguide on a diamond containing a 4.5 ppm density of NVs. The high-quality buried waveguides (type II geometry) are fabricated by femtosecond laser writing, and exhibit insertion loss below \SI{12}{dB} at \SI{635}{\nano\metre}. We characterize the spectrum and the spin coherence properties of NVs in the waveguide and pristine regions, showing that waveguide fabrication does not degrade their photoluminescence (PL) emission or spin coherence time. A fiber-waveguide-fiber configuration is then demonstrated to show enhanced sensing of a DC magnetic field. Thanks to the probing of the ensemble NVs along the whole waveguide device, this setup achieves at least an order of magnitude improvement in sensitivity compared to the traditional confocal configuration. Moreover, by using an excitation and collection path in a buried waveguide\cite{Eaton2019QuantumIrradiation}, the probed target could be potentially placed on the diamond surface without direct laser excitation\cite{doi:10.1073/pnas.1601513113}, therefore our study provides a non-invasive-way to future bio-sensing for samples vulnerable to optical illumination\cite{doi:10.1073/pnas.1601513113,Laissue2017AssessingImaging}.

\section{\label{Results and discussion} Results and discussion}
\begin{figure*}[ht]
\centering
\includegraphics[width=\textwidth]{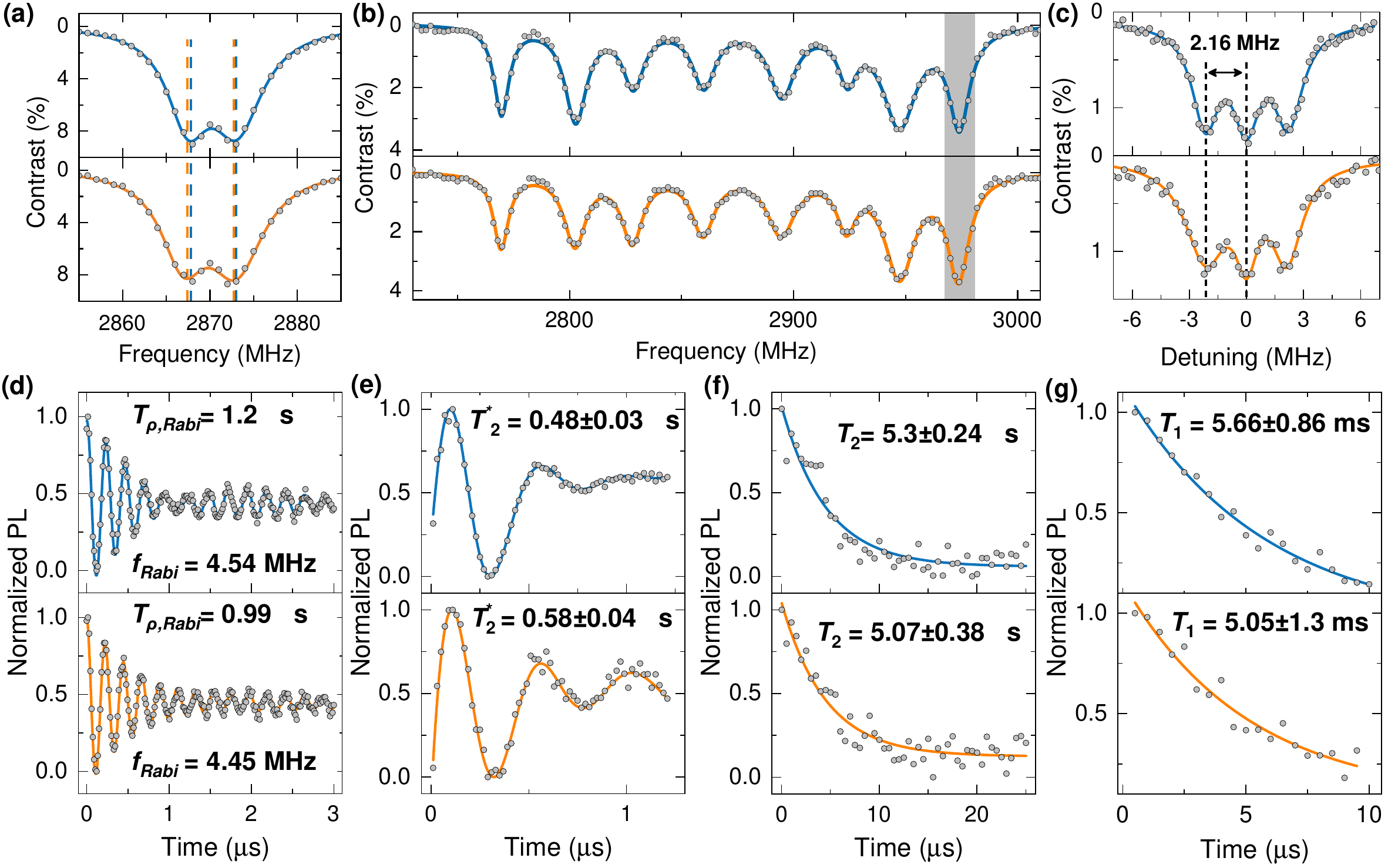}
\caption{Spin coherence properties of an ensemble of NVs in the pristine (upper plates) and waveguide (lower plates) regions. The grey points are the experimental data. The blue and orange lines are fitted curves. (a) Zero-field, CW ODMR. (b) CW ODMR with $\sim$\SI{5}{\milli \tesla} applied magnetic field where the data is fitted by the multiple Lorenz equation. (c) Pulsed ODMR for the highest resonance transition in (b). (d), (e) (f) and (g) are the Rabi oscillation, free-induced decay, Hahn echo, and spin longitudinal relaxometry measurements, respectively, where the experimental data is fitted by exponential decay equations\cite{Guo2024Laser-writtenDiamond}.}
\label{Fig2}
\end{figure*}

\subsection{Laser writing fabrication}

Our study uses optical waveguides in a commercial chemical vapor deposition (CVD) diamond (DNV-B14) from Element 6 which has a uniform and high concentration of NVs (4.5 ppm NVs with an inhomogeneous dephasing time $T_2^*$ of \SI{0.5}{\micro\second} at room temperature). As shown in Fig. \ref{Fig1}(a), optical waveguides consisting of pairs of parallel amorphous and graphitized modification lines were written by scanning laser pulses of \SI{515}{\nano\metre} wavelength, \SI{300}{fs} pulse duration, \SI{500}{kHz} repetition rate across the sample using a 100$\times$ 1.25 NA oil immersion Olympus objective, \SI{0.5}{mm/s} scan speed, \SI{18}{\micro\metre} depth and \SI{13}{\micro\metre} separation between optical modification lines\cite{Sotillo2017VisiblePulses}. There is no post-annealing after the laser writing fabrication. From left to right of Fig. \ref{Fig1}(b), we show four waveguides written with powers of 20, 30, 40, and 50 mW, where the overlaid mode profile in the cross-sectional image (lower plate) is the waveguide mode of a \SI{635}{\nano\metre} test laser from a beam-profiler. The mode parameters are detailed in TABLE \ref{Insertion loss}. As the laser fabrication power was increased from 20 mW to 50 mW, we found that the insertion loss decreased and that the optical modes were more tightly confined. The insertion loss includes diamond intrinsic absorbance (See UV-Via-NIR transmission spectrum of DNV-B14 diamond in supplemental materials (SM)), fiber-waveguide coupling loss, and waveguide propagation loss across the \SI{2.8}{mm} chip. The waveguide written with laser power of \SI{40}{mW} had a $\sim$\SI{5}{\micro\metre} mode field diameter (MFD) at \SI{635}{\nano\metre} and \SI{12}{dB} insertion loss is chosen for the further characterization of spectral properties, spin coherence properties and magnetic field sensing. 
\begin{table}[ht]
\centering
\caption{Insertion loss and mode field diameter at \SI{635}{\nano\metre} for the laser-written waveguide in diamond.}
\begin{tabular}{cccl}
\hline
Power (mW)
&Insertion loss (dB)& \ce{MFD_x}($\SI{}{\micro\metre}$)& \ce{MFD_y}($\SI{}{\micro\metre}$)\\
\hline
20
&14.2&  5.5& 6.0\\
30
&12.1&  4.3& 6.2\\
40
&11.6&  4.8& 5.5\\
 50&11.6& 4.2& 4.9\\
 \hline
\end{tabular}
  \label{Insertion loss}
  \end{table}
  
\subsection{PL study}

We used a custom optically detected magnetic resonance (ODMR) confocal setup to characterize the waveguide's spectrum and spin coherence properties. The setup is detailed in the SM. In Fig. \ref{Fig1}(c), measuring the sample from its facet edge, end on, we directly map the waveguide cross-section, which resolves three regions labeled as  1-pristine, 2-sidewall, and 3-waveguide regions. In Fig. \ref{Fig1}(d), the spectra of 1-pristine and 3-waveguide areas feature similar spectral properties with a clear \SI{637}{\nano\metre} zero phonon line and broad phonon sideband extending to almost \SI{800}{\nano\metre}. The PL intensity was reduced in the sidewall areas where the laser modification lines have converted the diamond to graphitized and amorphous carbon\cite{Eaton2019QuantumIrradiation}. Notably, the PL intensity and spectra in pristine and waveguide regions are identical indicating the laser writing process has not degraded the NVs.

\subsection{Spin coherence characterization}

Spin coherence proprieties are core to quantum sensing, enabling the detection of weak perturbations to the spin's environment. As a result of the NVs' efficient spin-selective transition, the spin coherence properties of its \textbf{\textit{S}}=1 triplet ground state can be easily characterized via ODMR techniques\cite{Rondin2014MagnetometryDiamond}. 

In the absence of the magnetic field, strain, and electric field, the ground state's $m_s=\pm1$ sublevels are degenerate and separated by the axial zero-field splitting parameter $D$=\SI{2870}{MHz} from the $m_s=0$ sublevel which arises from the electron spin-spin interactions\cite{Rondin2014MagnetometryDiamond}. These energy differences between the ground state sublevels can then be read out by recording NVs PL intensity while scanning the microwave frequency near resonance. This frequency domain zero-field ODMR contains information on microscopic local environment coupling with the NVs electron spin\cite{Mittiga2018ImagingDiamond}. In Fig. \ref{Fig2}(a), the zero-field continuous-wave (CW) ODMR in the pristine and waveguide regions exhibit a $\sim$ \SI{3}{MHz} transverse zero-field splitting parameter of $E$ due to the local stain and electric field from the diamond crystal\cite{Rondin2014MagnetometryDiamond}. Additionally, the broader linewidth ($w_1$=\SI{6.2}{MHz} and $w_2$=\SI{7.8}{MHz}) of the $m_s=\pm1$ resonance peaks in the waveguide region indicates the increased non-hydrostatic strain induced by laser-written fabrication process\cite{Eaton2019QuantumIrradiation,Mittiga2018ImagingDiamond}, compared to the $w_1$=\SI{5.7}{MHz} and $w_2$=\SI{7.2}{MHz} observed in pristine region. 
A $\sim$\SI{5}{mT} magnetic field is applied to lift the degeneracy of the $m_{s}$=$\pm 1$ transitions along the four NV orientations in Fig. \ref{Fig2}(b). The highest frequency transition is further investigated with pulsed ODMR in Fig. \ref{Fig2}(c), where the typical \SI{2.16}{\mega\hertz} \ce{^{14}N} hyperfine splitting is resolved. 

Electron spin Rabi oscillations, free-induction decay, Hahn-echo, and spin-lattice relaxometry of ensemble NVs in pristine and waveguide regions were measured with standard protocols\cite{Guo2024Laser-writtenDiamond} in Fig. \ref{Fig2}(d-g). The NV ensembles in both areas are shown to have comparable spin coherence times of  Rabi oscillation decoherence time $T_{\rho,}\textsubscript{Rabi}\sim$\SI{1}{\micro\second}, inhomogeneous dephasing time $T_2^*$$\sim$\SI{0.5}{\micro\second}, spin transverse relaxation time $T_2$$\sim$\SI{5}{\micro\second}, and longitudinal relaxation times $T_1$$\sim$\SI{5}{m\second}. This implies that laser writing fabrication does not degrade the spin coherence properties of the ensemble\cite{Guo2024Laser-writtenDiamond}. This should be contrasted with other fabrication methods that have been used to create photonic structures in diamond, such as plasma etching \cite{Wang2022Self-alignedPrecision,Babinec2010ASource}, and focused-ion beam\cite{Hadden2010StronglyLenses}, which have been shown to have a detrimental effect on these parameters\cite{Wang2022Self-alignedPrecision,Barry2020SensitivityMagnetometry}.   

\begin{figure*}
\centering
\includegraphics[width=15cm]{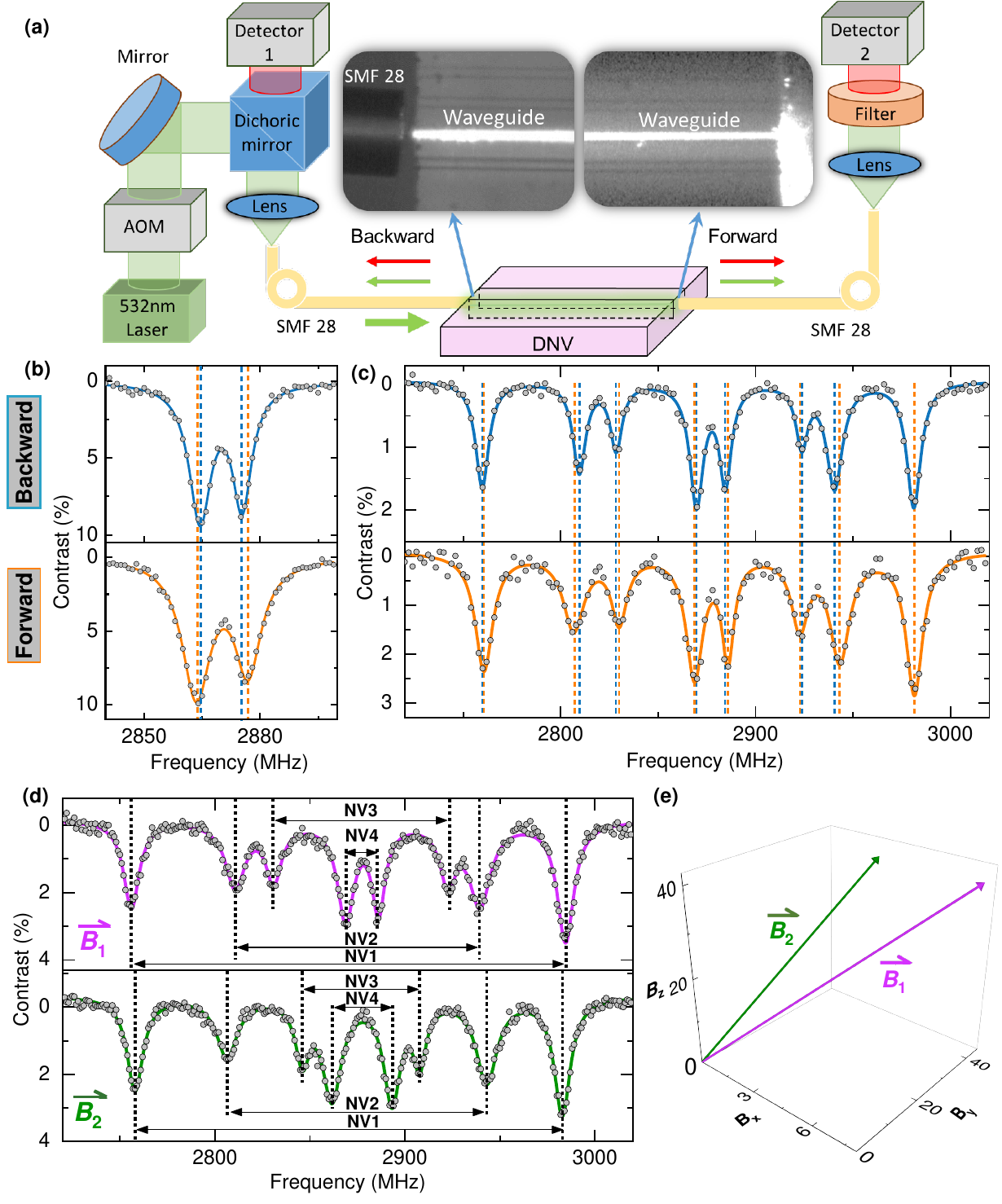}
\caption{Enhanced quantum sensing via fiber-waveguide-fiber configuration. (a) Schematic diagram of fiber-waveguide-fiber configuration where the insert two microscopy images show backward and forward travel light in the waveguide. (b) Zero-field ODMR recorded by backward (upper plates) and forward (lower plates) PL. (c) ODMR with applied magnetic field are recorded by backward (upper plates) and forward PL (lower plates) where the vertical blue and orange dash lines represent the resonance frequencies from forward and backward traveling ODMR curve fitting. (d) Magnetic field sensing via the forward traveling waveguide ODMR, where vertical black lines are the resonance frequencies responding to probed magnet fields $\overset{\large\rightharpoonup}{\small{B_1}}$ and $\overset{\large\rightharpoonup}{\small{B_2}}$ . (e) Magnetic field vector into Cartesian coordinates inferred from (d).}
\label{Fig3}
\end{figure*}

\subsection{Magnetometry in a fiber-waveguide-fiber configuration}

Compared to the conventional free space confocal setup for limited numbers of NVs, a fiber-waveguide-fiber configuration was used for high-density ensemble waveguide integrated NVs in Fig. \ref{Fig3}(a). Two SMF-28 single-mode fibers were used to couple to opposing facets to probe NVs along the entire waveguide efficiently. The input fiber delivers green laser excitation and collects the backward traveling red fluorescence from the NVs in the waveguide, which is filtered with a dichroic splitter and directed to a first detector. The second fiber solely collects the forward-traveling red fluorescence from the waveguide and directs it to a second detector. Detailed information on the fiber-waveguide-fiber setup is available in the SM.

Fig. \ref{Fig3}(b) shows the zero-field ODMR recorded using the backward and forward traveling PL emission from the chip. Compared to the confocal zero-field ODMR in Fig. \ref{Fig2}(a), the larger transverse zero-field splitting parameters $E$ (\SI{6.7}{MHz} for backward and \SI{5.4}{MHz} for forward ODMR) are observed in the fiber-waveguide-fiber configuration in Fig. \ref{Fig3}(b). This implies the average effect of the strain-induced and electric field-induced splitting for the ensemble NVs along the whole waveguide region\cite{Hoese2021IntegratedArrays}. Moreover, the 10\% ODMR contrast suggests an excellent ODMR response potentially leading to high sensitivity. Additional laser power and microwave power-dependent ODMR are included in the SM. We also observed a comparable ODMR response with the applied magnetic field in Fig. \ref{Fig3}(c) using the forward and backward traveling fluorescence from the chip. Eight resonance peaks resulting from the four different NVs orientations were clearly resolved. The contrast of the highest frequency peak in the ODMR spectrum is over 3\%, comparable to the confocal ODMR results. Meanwhile, there are also some resonance frequency shifts between forward and backward ODMR. This potentially could be used to determine the magnetic field gradient along the waveguide length of \SI{2.8}{\milli \metre}, but is beyond the scope of this work. 

By solely taking forward traveling PL, in Fig. \ref{Fig3}(d), applied magnet fields $\overset{\large\rightharpoonup}{\small{B_1}}$ and $\overset{\large\rightharpoonup}{\small{B_2}}$ are probed by tracking four distinct Zeeman split pairs of peaks originating from four different <111> NV$_i$ orientations. The magnetic field  projection $B_i$ along the NV$_i$  are calculated by the Zeeman effect in the ground state via the equation\cite{Rondin2014MagnetometryDiamond},

\begin{equation}
    \nu_{\pm}(B_i)= \textit{D}\pm\sqrt{(\frac{g\mu_B}{\textit{h}}B_i)^2+E^2}
\end{equation}

\noindent where $\nu_{\pm}(B_i)$ is the ODMR resonance frequency for different Zeeman splitting from four different NV orientations, the $g\sim$ 2\textit{.}0 is the Landé $g$-factor, $\mu_{B}$ is the Bohr magneton, and \textit{h} is the Planck constant. As shown in Fig. \ref{Fig3}(e), we inferred the magnetic field vectors $\overset{\large\rightharpoonup}{\small{B_1}}$ and $\overset{\large\rightharpoonup}{\small{B_2}}$ in Fig. \ref{Fig3}(d) via simple geometric arguments to transform the tetrahedral directions into Cartesian coordinates\cite{maertz2010vector}. We note the magnetic field vector in lab coordinates can be defined by the excitation polarization plot for four differently oriented NVs\cite{weggler2020determination,chen2020calibration}. 

\subsection{Enhanced sensitivity}

The intrinsic sensitivity is not only dependent on the strong response to the target signal but also on avoiding interactions with undesirable noise\cite{Degen2017QuantumSensing}. Generally, the photon-shot-noise-limited DC ($\eta\textsubscript{dc}$)  and AC ($\eta\textsubscript{ac}$) sensitivity can be used to quantify the sensitivity, and are given by \cite{Rondin2014MagnetometryDiamond},
\begin{equation}
\eta\textsubscript{dc}\sim\frac{\hbar}{g\mu_B}\frac{1}{\Lambda\sqrt{Ct_L}}\times\frac{1}{\sqrt{T_2^*}},
\label{eq:n_dc}
\end{equation}
\begin{equation}
\eta\textsubscript{ac}=\eta\textsubscript{dc}\sqrt{\frac{T_2^*}{T_2}},
\label{eq:n_ac}
\end{equation}
respectively, where $\hbar$ is the reduced Planck constant, and $t_{L}\sim$\SI{0.5}{\micro\second} is the readout duration time. $T_2^* $ is $\sim$\SI{0.5}{\micro\second} and $T_2$ is $\sim$\SI{5}{\micro\second} where the spin coherence properties of the WGINVs are consistent with that of native NVs in pristine regions. $\Lambda\sim3\%$ is ODMR contrast. $C$ is the total detected PL rate relying on the excitation power and experiment configuration. For a conventional top-down confocal configuration with NA=0.9 objective, the saturation PL rate of \SI{362.4}{GHz} and saturation power of 11.7 mW is achieved, resulting in $\eta\textsubscript{dc}$\SI{627}{\pico T\cdot\hertz^{-1/2}} and $\eta\textsubscript{ac}$\SI{198}{\pico T\cdot\hertz^{-1/2}}\cite{Rondin2014MagnetometryDiamond}. The detailed power-dependent PL data in SM.





In the fibre-waveguide-fiber configuation, the saturation PL rate is affected by a number of factors such as the (i) enlarged mode area in the waveguide relative to the confocal spot\cite{Guo2024Laser-writtenDiamond}, which increases the number of NVs probed in the unit optical plane and the saturation laser power, by a factor of $\sim$100. (ii) There is an increased optical depth of NVs probed by the beam as it propagates along the waveguide, relative to the Rayleigh range of excitation volume in the confocal configuration\cite{Guo2024Laser-writtenDiamond}. (iii) The waveguides have a reduced numerical aperture, arising from their weak confinement, of the order of NA=0.012 \cite{Hadden2018IntegratedWriting} compared to a NA in the confocal system of NA$\sim$0.4 inside the diamond. (iv) There is a coupling loss due to mode mismatch between the fiber and waveguide.


We experimentally assess how these effects combine to scale the PL rate in the fiber-waveguide-fiber configuration. A laser power of  \SI{1.5}{mW}, much less than saturation power, was used to evaluate the PL emission in fiber-waveguide-fiber configuration. We observed \SI{40}{GHz} PL rate in the backward-traveling fluorescence and \SI{4}{GHz} PL rate in the forward direction, which is comparable to the PL emission of \SI{42}{GHz} in the confocal configuration under same laser power. This indicates that the fiber-waveguide-fiber configuration achieves comparable PL emission to the confocal setup by probing a larger NV volume, despite its 100 times enlarged excitation area reducing excitation laser power density by 100\cite{Guo2024Laser-writtenDiamond}. Therefore, we can estimate that fiber-waveguide-fiber configuration increases the saturation power and saturation PL rate by a factor of 100, offering a 10-fold improvement of sensitivity of $\eta\textsubscript{dc}$= \SI{63}{\pico T\cdot\hertz^{-1/2}} and $\eta\textsubscript{ac}=$ \SI{20}{\pico T\cdot\hertz^{-1/2}} compared to confocal configuration.



This level of sensitivity makes this waveguide geometry quantum sensor competitive with other platforms in the $\sim$ \SI{10}{\micro m} length scale such as superconducting quantum interference devices and Hall sensors\cite{degen2008microscopy}. Furthermore, the possibility of integrating microfluidic channels\cite{Eaton2019QuantumIrradiation} for ion transport and neuron imaging\cite{doi:10.1073/pnas.1601513113} is attractive for this technology.

\section{Conclusion}

We have introduced an approach for waveguide-enhanced quantum sensing using a high-quality buried waveguide in diamond chip with 4.5 ppm NV density. We show that femtosecond laser writing does not change the density, spectrum, and spin coherence properties of the NVs, leading to a high sensitivity and increased optical depth with ten micrometers of resolution. In a fiber-waveguide-fiber setup, we obtained robust ODMR response from which we inferred the magnetic field magnitude and direction. The increased NV density in this chip, combined with the advantages of using a fiber-waveguide coupled system leads to $\eta\textsubscript{dc}$=\SI{63}{\pico T\cdot\hertz^{-1/2}} and $\eta\textsubscript{ac}$=\SI{20}{\pico T\cdot\hertz^{-1/2}}. Other than the improved sensitivity, WGINVs in DNV-B14 are significantly more reproducible compared to the WGINVs in high-pressure high-temperature diamond in our previous report\cite{Guo2024Laser-writtenDiamond}, thanks to a more uniform NV density in the CVD produced sample. Future work will focus on applying this technology to electric field sensing or bio-sensing. Alternative photonic structures with integrated NVs could be tailored for different applications by leveraging the adaptability of the femtosecond laser direct-write process.

\section*{Author Contributions}

The corresponding author identified the following author contributions, using the CRediT Contributor Roles Taxonomy standard:

YG: Conceptualization, Methodology, Software, Data curation, 
Formal Analysis, Investigation, Writing - Original Draft, Writing - Review \& Editing.
GC: Resources, Writing - Review \& Editing.
VKK: Resources, Writing - Review \& Editing.
ANG: Resources, Writing - Review \& Editing.
ANV: Resources, Writing - Review \& Editing.
RR: Writing - Review \& Editing, Funding Acquisition.
CS: Resources, Writing - Review \& Editing, Supervision.
PEB: Resources, Writing - Review \& Editing, Supervision.
JPH: Conceptualization, Writing - Review \& Editing, Supervision, Funding Acquisition.
AJB: Conceptualization, Resources, Writing - Review \& Editing, Supervision, Funding Acquisition.
SME: Conceptualization, Resources, Writing - Review \& Editing, Supervision, Funding Acquisition.

\section*{Acknowledgments}

The authors acknowledge financial support provided by EPSRC via Grant No. EP/T017813/1 and EP/X03982X/1 and the European Union's H2020 Marie Curie ITN project LasIonDef (GA No. 956387). IFN-CNR is thankful for support from the projects QuantDia (FISR2019-05178) and PNRR PE0000023 NQSTI funded by MUR (Ministero dell'Università e della Ricerca). S. M. Eaton is grateful to the Department of Physics at Politecnico di Milano for access to the FELICE laboratory for crucial characterization experiments. NTU thanks the Singapore National Research Foundation (NRF2021-QEP2-01-P01) and NRF-MSG-NCAIP. The authors thank Sam Bishop's help and discussion on the initial optical setup.  

\section*{Disclosures}
The authors declare no conflicts of interest.

\section*{Data Availability Statement}
Data supporting the findings of this study are available in the Cardiff University Research Portal at http://doi.org/xx.xxxx.

\bibliography{references}
\onecolumngrid
\clearpage
\begin{center}
\textbf{\large Supplemental Materials: Enhanced quantum magnetometry with a laser-written integrated photonic diamond chip}
\end{center}

\setcounter{equation}{0}
\setcounter{figure}{0}
\setcounter{table}{0}
\setcounter{page}{1}
\makeatletter
\renewcommand{\theequation}{S\arabic{equation}}
\renewcommand{\thefigure}{S\arabic{figure}}
\renewcommand{\bibnumfmt}[1]{[S#1]}
\setcounter{section}{0}

\section{\label{Confocal ODMR setup} Confocal ODMR setup}

\begin{figure}[ht]
    \centering
    \includegraphics[width=0.4\linewidth]{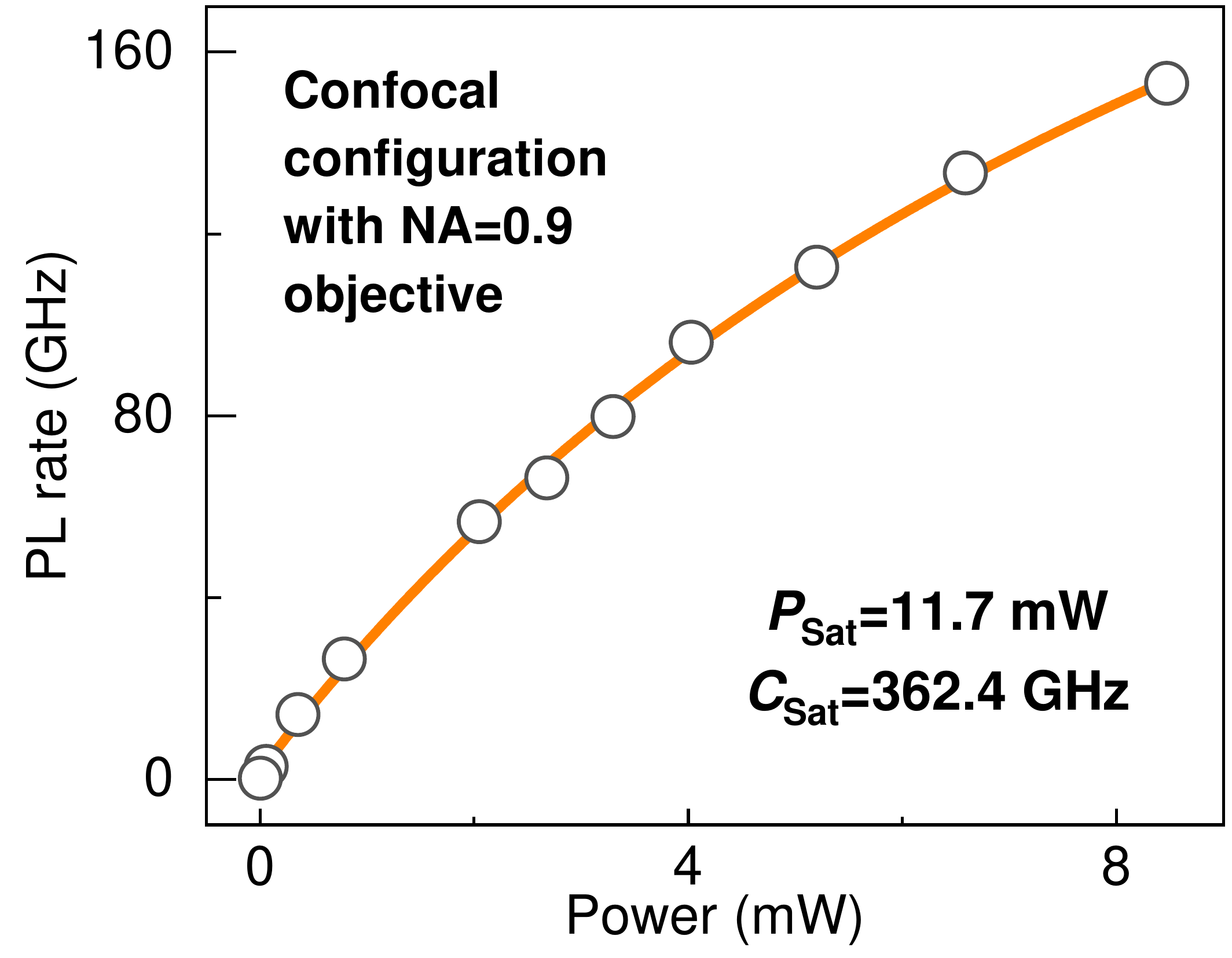}
    \caption{Power-dependent saturation behavior for purple diamond under confocal configuration.}
    \label{fig:S1}
\end{figure}

A CW \SI{532}{\nano m} crystal laser was modulated by an acoustic-optic modulator (ISOMET 553F-2) with < \SI{10}{\nano s} rise and fall time. A 2-axis Galvo mirror (GVS002) and 100 $\times$ Nikon objective with NA=0.9 were integrated into a 4f imaging system for 2D x-y scanning. Depth scanning (z) was implemented by a motorized sample stage. The PL was optically filtered by the dichroic mirror, \SI{532}{\nano m} long-pass filter, and \SI{650}{\nano m} long-pass filter, before detection on SPCM-AQRH silicon avalanche photodiodes (Excelitas) or a spectrometer with a silicon CCD. The optional ND filter is also used to keep the PL rate within the APD's linear response range (\SI{2}{\mega\hertz}) for the power-dependent PL saturation measurement in Fig. \ref{fig:S1}.  The microwave (MW) field is generated by an E4438B MW source, modulated by a RF switch (ZASWA-2-50DRA+) and amplified by a MW amplifier (ZHL-42W+), and eventually transmitted to the sample by a patch antenna. The Rabi oscillation, free induction decay, Hahn echo, and $T_{1}$ measurements are implemented by the standard protocol.

\section{\label{UV-via-NIR transmission spectrum for pristine DNV-B14 diamond} UV-Via-NIR transmission spectrum for pristine DNV-B14 diamond}

\begin{figure}[ht]
    \includegraphics[width=8cm]{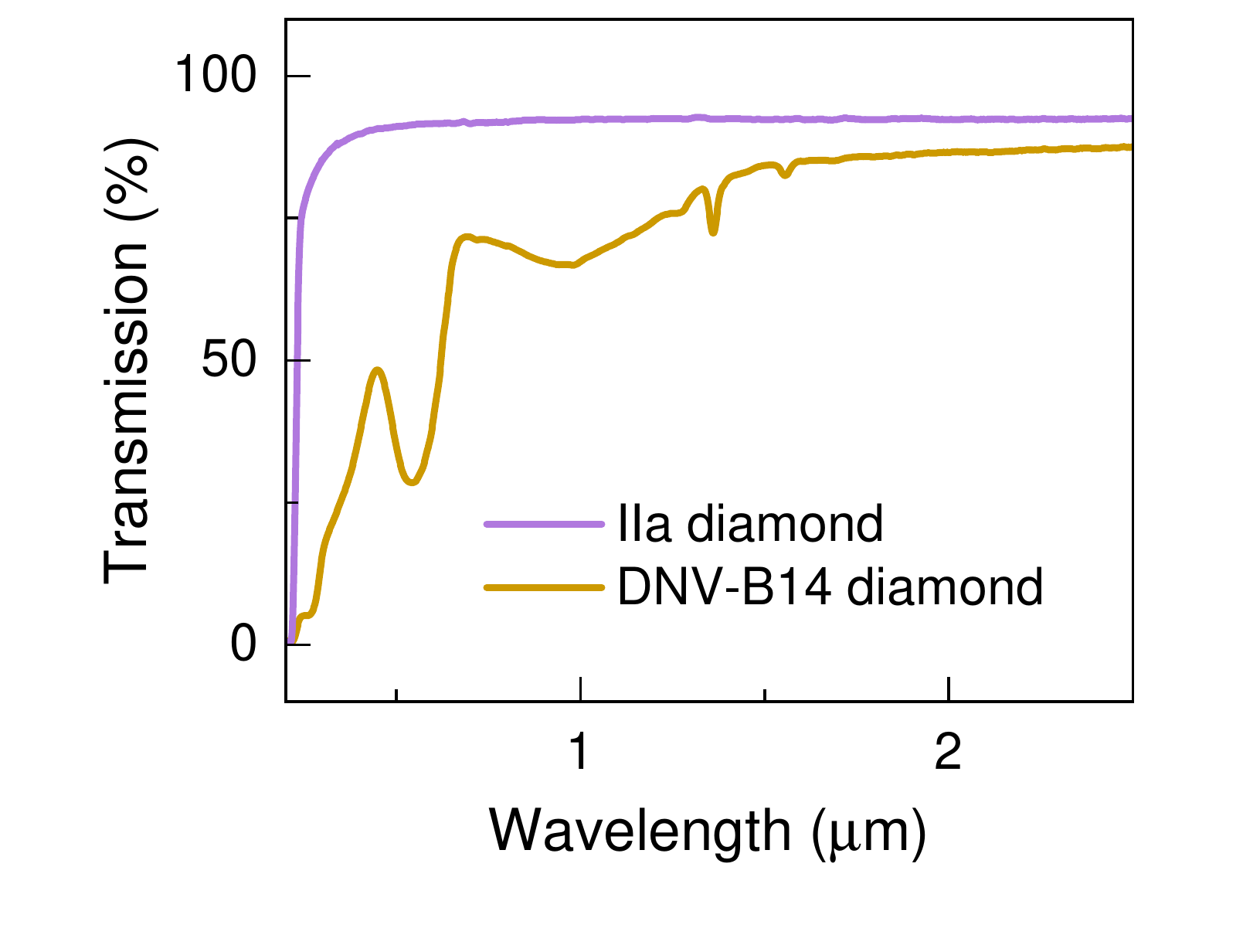}
    \label{Fig:S2}
 \caption{UV-Vis-NIR transmission spectra for pristine DNV-B14 diamond and IIa diamond}
\end{figure}

The UV-Via-NIR spectra of the DNV-B14 diamond and IIa diamond were measured using an MSV-5200 Microspectrophotometer. In Fig. \ref{Fig:S2}, compared to high pure IIa diamond, we observed non-negligible loss in DNV-B14 diamond.

\section{\label{Fiber-waveguide-fiber configuration} Fiber-waveguide-fiber configuration}

In the fiber-waveguide-fiber configuration, the input laser is coupled into a fiber polarization controller for polarization tuning, then connected to SMF-28 Ultra single-mode fiber, coupled into waveguide in diamond. The green laser would excite the NVs along the waveguide which transmit the PL emission and the green laser along the waveguide. Another SMF-28 Ultra single-mode fiber, as output fiber, would extract the waveguide mode of PL emission and green laser into the fiber space again where the PL emission and green laser are separated by the \SI{532}{\nano\metre} and \SI{650}{\nano\metre} long pass filter in free space. The APD and optional ND filters are used to read out the signals. 

\section{\label{Power dependent zero field ODMR} Power dependent zero field ODMR}

\begin{figure}[ht]
    \includegraphics[width=17 cm]{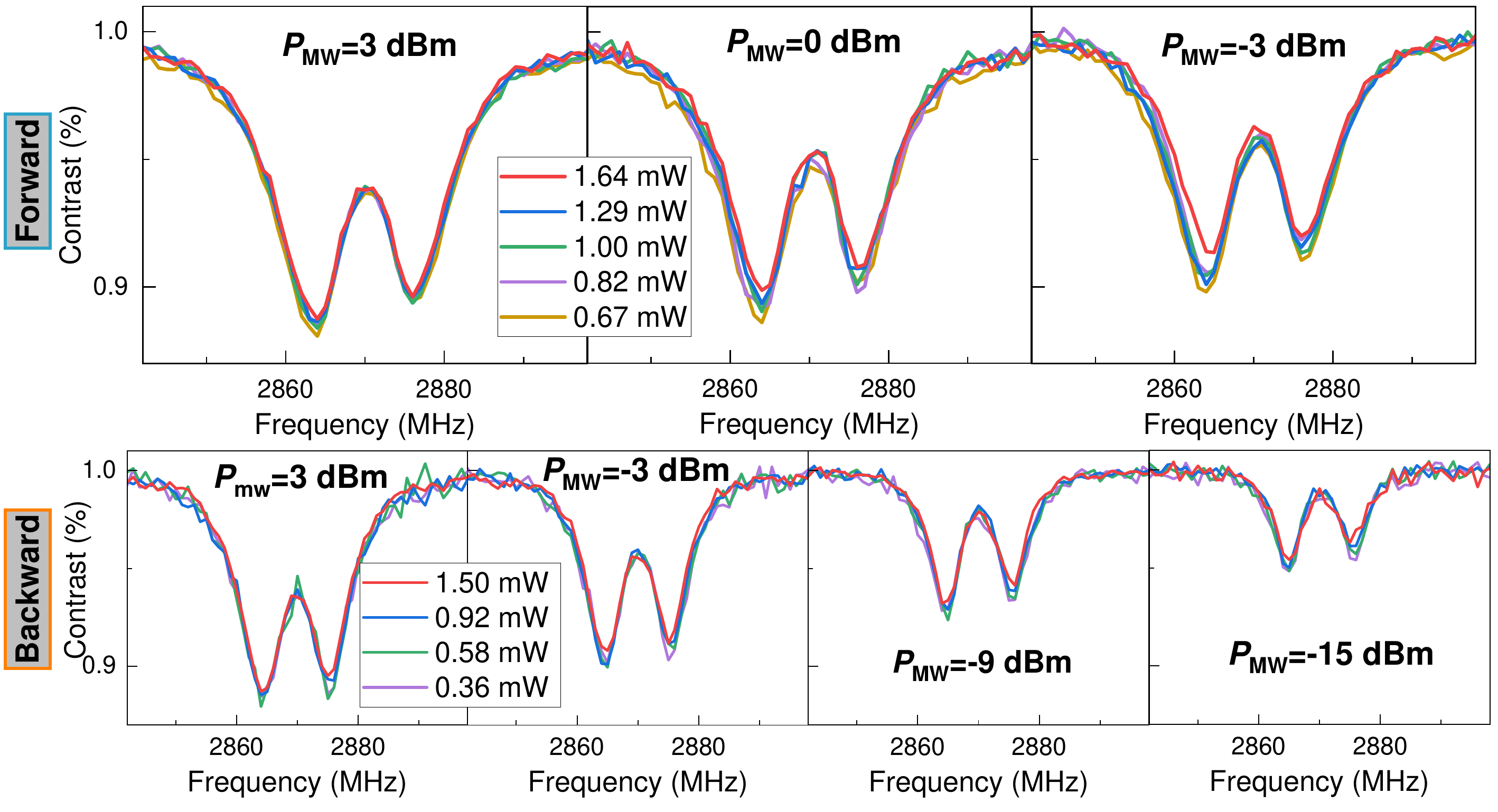}
    \label{Fig:S3}
    \caption{The laser power and microwave power-dependent ODMR, where $P\textsubscript{MW}$ is the MW power from the direct output of E4438B MW source}
\end{figure}

In confocal microscopy, the laser power and MW power greatly impact the ODMR shape. This is typical due to the non-negligible power broadening and dynamics completion between laser polarization and MW manipulation in the ground state. In the fiber-waveguide-fiber configuration, from Fig.\ref{Fig:S3}, the laser excitation power appears to play less role than the microwave in the ODMR spectrum recorded by forward and backward traveling PL. This behavior might be because the excited laser powers are far less than the saturation power for fiber-waveguide-fiber configuration. 

\end{document}